\documentclass[12pt,preprint]{aastex}

\begin{document}
\title{Observation of the Molecular Zeeman Effect in the G band}

\author{A. Asensio Ramos\altaffilmark{1}, J. Trujillo Bueno\altaffilmark{1,2},
M.
Bianda\altaffilmark{3,4},
R. Manso Sainz\altaffilmark{1,5}, H. Uitenbroek\altaffilmark{6}}
\altaffiltext{1}{Instituto de Astrof\'{\i}sica de Canarias, V\'{\i}a L\'actea s/n, E-38205, La Laguna,
Tenerife, Spain}
\altaffiltext{2}{Consejo Superior de Investigaciones Cient\'{\i}ficas (Spain)}
\altaffiltext{3}{Istituto Ricerche Solari Locarno, Via Patocchi, CH-6605 Locarno-Monti, Switzerland}
\altaffiltext{4}{Institute of Astronomy, ETH Zentrum, CH-8092 Z\"urich, Switzerland}
\altaffiltext{5}{High Altitude Observatory, National Center for Atmospheric Research\altaffilmark{7}, P. O. Box 3000,
Boulder CO 80307-3000}
\altaffiltext{6}{National Solar Observatory, P. O. Box 62, Sunspot, NM 88349}
\altaffiltext{7}{The National Center for Atmospheric Research is sponsored by the National Science Foundation}
\begin{abstract}

Here we report on the first observational investigation
of the Zeeman effect in the G band around 4305 \AA. Our
spectropolarimetric observations of sunspots
with the Z\"urich Imaging Polarimeter at the IRSOL facility
confirm our previous
theoretical prediction that the molecular Zeeman effect
produces measurable circular polarization signatures
in several CH lines that are
not overlapped with atomic transitions. We also find both
circular and linear polarization signals produced by
atomic lines whose wavelengths lie in the G band
spectral region. Together, such molecular and atomic lines are
potentially important for empirical investigations
on solar and stellar magnetism. For instance,
a comparison between observed and calculated
Stokes profiles suggests that the thermodynamical and/or magnetic properties of the photospheric regions
of sunspot umbrae are horizontally structured with a component that might be associated with umbral dots.

\end{abstract}


\section{Introduction}

The Fraunhofer G band around 4305 \AA\
includes significant contributions from vibro-rotational
transitions of the
${\rm A}^2{\Delta}-{\rm X}^2{\Pi}$ electronic system of the CH molecule.
This band has been important in high spatial resolution studies of solar
surface magnetism because photospheric bright points
are seen with very high contrast when observed in the G band (see S\'anchez Almeida et al. 2001 and references therein; see also
the recent G band images of Scharmer et al. 2002
and of S\'anchez Almeida et al. 2004 which show
bright points around sunspots and in the internetwork quiet Sun, respectively).
For this reason, theoretical and observational investigations of the Zeeman effect in the G band are of great scientific interest.

In a recent paper (see Uitenbroek et al. 2004) we
have investigated theoretically the
Zeeman effect in the G band, including radiative
transfer simulations of the emergent Stokes profiles in `quiet' sun models.
In spite of the fact that this spectral region contains a multitude of
overlapping atomic and molecular lines we could find
several isolated groups of CH lines that were predicted to produce
measurable circular polarization signals in the presence of magnetic fields.
In particular, in one wavelength location near 4304 \AA\ the overlap
of several magnetically sensitive and non-sensitive CH lines
was predicted to produce a single-lobed Stokes $V$ profile that
could be of great interest for future high spatial-resolution narrow-band
polarimetric imaging of the solar surface.

The main aim of this letter is to report on the first
observational study of the Zeeman effect in the G band,
but including also radiative transfer calculations in a
semi-empirical model of sunspot atmospheres.
Our spectropolarimetric observations of sunspots
fully confirm the above-mentioned theoretical
prediction. As we shall see below, a preliminary comparison
of the observed and calculated Stokes profiles
demonstrates that the Zeeman effect in the G band
offers a new diagnostic window for empirical investigations
on solar and stellar magnetism.

\section{Theoretical prediction of the G band polarization in sunspots}

Prior to reporting on our spectropolarimetric observations
it is useful to show first the emergent Stokes profiles
that we expect for the atmospheres of sunspot umbrae.
To this end, we have used the cool semi-empirical
model of Collados et al.
(1994), but assuming a
vertical magnetic field of constant
strength. This is sufficient for our demonstrative purposes ---that is, for
showing that the \emph{shapes} of the computed $V/I$ profiles agree
with those we have observed in sunspots and for highlighting the
diagnostic interest of the Zeeman effect in the G band.
The vector radiative transfer equation
for the emergent Stokes profiles has been
solved by applying the
quasi-parabolic DELO method (Trujillo Bueno 2003), in a way similar to
that described by Uitenbroek et al. (2004).

Figure \ref{fig_syn} shows the calculated $V/I$ profiles
in two spectral regions of the G band assuming a vertical magnetic field of 2000 G.
The left and right panels
correspond to the 4304 \AA\ and 4312 \AA\ spectral regions, respectively.
The chosen line of sight is given by $\mu=\cos \theta=0.95$,
where $\theta$ is the heliocentric angle.
It is important to advance here that the observed Stokes profiles described
in Section 3 are affected by a stray-light contamination
of $\sim$4\% , which we have determined by comparing
the intensity profiles of Kitt Peak's atlas
observed with the Fourier Transform Spectrometer (FTS) with
our own observations at the solar disk center.
For this reason, in Fig. \ref{fig_syn} we have also shown
the artificial reduction in the {\em amplitude} of the computed $V/I$ profiles
that results when a 4\% contribution of the intensity profiles
obtained in the quiet Sun model
FAL-C (Fontenla et al 1993) is added
to the calculated Stokes $I$ profiles in the umbra model.
The solid line shows the emergent Stokes profiles calculated by including
both the atomic and
CH lines, while the
dashed line represents the emergent Stokes profiles obtained by including only
the CH lines. The dashed-dotted line represents the emergent Stokes profiles
calculated without including any stray-light contamination.

The wavelengths and oscillator strengths for the CH lines have been obtained
from J\o rgensen's et al (1996) linelist while
those for the atomic lines have been obtained from CD-ROM 1 by
Kurucz\footnote{\texttt{http://kurucz.harvard.edu}}.
In both spectral regions we have $V/I$ features dominated by CH lines
only. For example,
the one-lobe $V/I$ profile at 4304 \AA\ results
from the blend between the ${\rm R}_{1e}(1.5)$
and ${\rm R}_{1f}(1.5)$ lines of the $v=0-0$ band, which
gives a $\Lambda$-doublet with effective Land\'e factor
$\bar{g}=0.8833$. The
$V/I$ profile at 4313.7 \AA\ is
produced by the blend of the CH lines ${\rm Q}_{11fe}(3.5)$
and ${\rm Q}_{11ef}(3.5)$. Finally, the $V/I$ feature at 4312 \AA\
is produced by many overlapping CH lines, but mainly
by the ${\rm P}_{12ff}(5.5)$, ${\rm R}_{21ff}(4.5)$, ${\rm Q}_{22fe}(4.5)$
and ${\rm P}_{12ee}(4.5)$ from the $v=0-0$ band and by the
${\rm R}_{21ff}(1.5)$, ${\rm P}_{12ff}(4.5)$, ${\rm R}_{21ee}(1.5)$
and ${\rm R}_{21ff}(2.5)$ from the $v=1-1$ band. Note that the CH
lines that produce clean $V/I$ signals result from transitions
between levels with small $J$-values (i.e., with $J{\le}5.5$).

\section{Observations of the G band polarization in sunspots}

The spectropolarimetric observations were carried out on 2003 August
30 using the UV version of the Z\"urich Imaging Polarimeter (ZIMPOL; see Povel
2001) attached to the Gregory Coud\'e Telescope
of the Istituto Ricerche Solari Locarno
(IRSOL). In order to facilitate an irrefutable
observational proof of the predicted Stokes $V/I$ profiles
we selected a bipolar sunspot group (NOAA 0447), which was located
at $\mu=0.95$. We observed simultaneously the
two sunspots of opposite magnetic polarity.
To this end, a Dove prism located
after the analyzer (modulator + linear polarizer)
allowed us to rotate the solar image
in order to get the image of the umbrae of the two main
spots on the spectrograph
slit. The direction of the slit formed an angle
of $55^\circ$ with respect to the closest limb. The
slit width was 80 $\mu$m, which
corresponds to 0.7\arcsec\  on the solar surface. The spatial and spectral
extensions covered by the ZIMPOL CCD were 160\arcsec\ and
3.1 \AA, respectively.

The ZIMPOL version used for these observations has
one piezoelastic modulator
(PEM), which allows to measure simultaneously
the intensity $I$, and the fractional Stokes parameters $Q/I$ and $V/I$ for linear and
circular polarization, respectively. Due
to the high modulation rate
(42 kHz which is much higher than the typical
seeing frequencies of the
order of hundreds of Hertzs) the seeing-induced crosstalk
is insignificant
and the noise level in the polarization signals is
established only by
photon statistics. ZIMPOL is a one beam
system and the same pixel of
the CCD is used to measure all Stokes components.
Therefore, no flatfield technique
is needed to correct the polarization images.
A calibration is performed by inserting known
amounts of polarization in
front of the analyzer (PEM +
linear polarizer).
The dark current table correction is also applied.
Our measurements were
performed by adding 60
registrations, each of them taken with an integration time of 5 seconds.
The reduced data are affected by
instrumental polarization caused
by the two folding mirrors inside
the Gregory Coud\'e Telescope (for details
see Gandorfer et al. 1997). The ensuing effects
(i.e., a shift of the zero polarization level
smaller than 1\% and a 2\% crosstalk from $V/I$ to $Q/I$)
were quite small and constant over the day
and could be easily corrected.
In order to increase
the signal-to-noise ratio of the $V/I$
profiles shown in the following two figures, we averaged over 4 pixels along
the spatial direction, which
corresponds to 4.5\arcsec\ inside the sunspot umbrae.
No other data reduction procedures like
smoothing or filtering were applied.

\begin{figure}
\plottwo{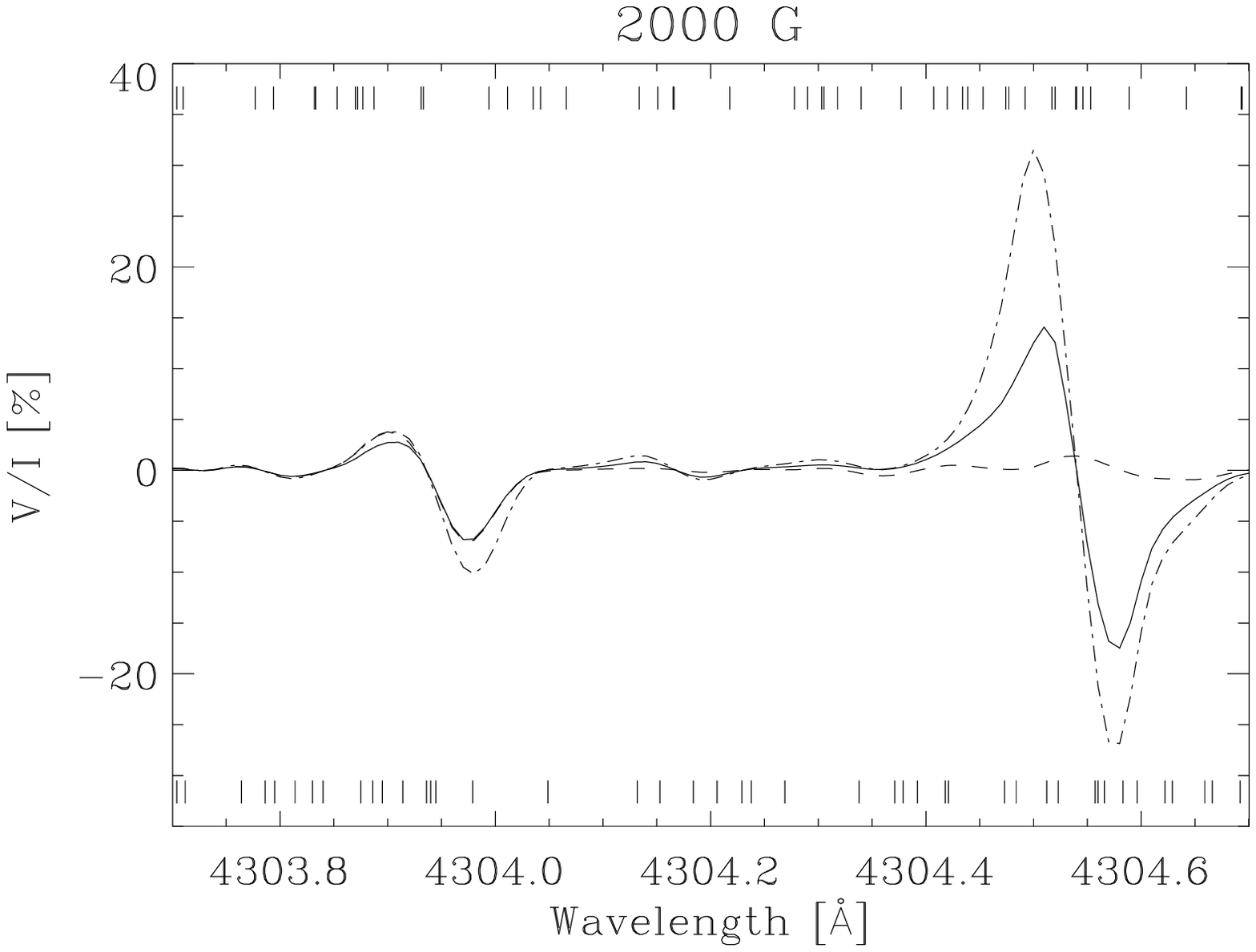}{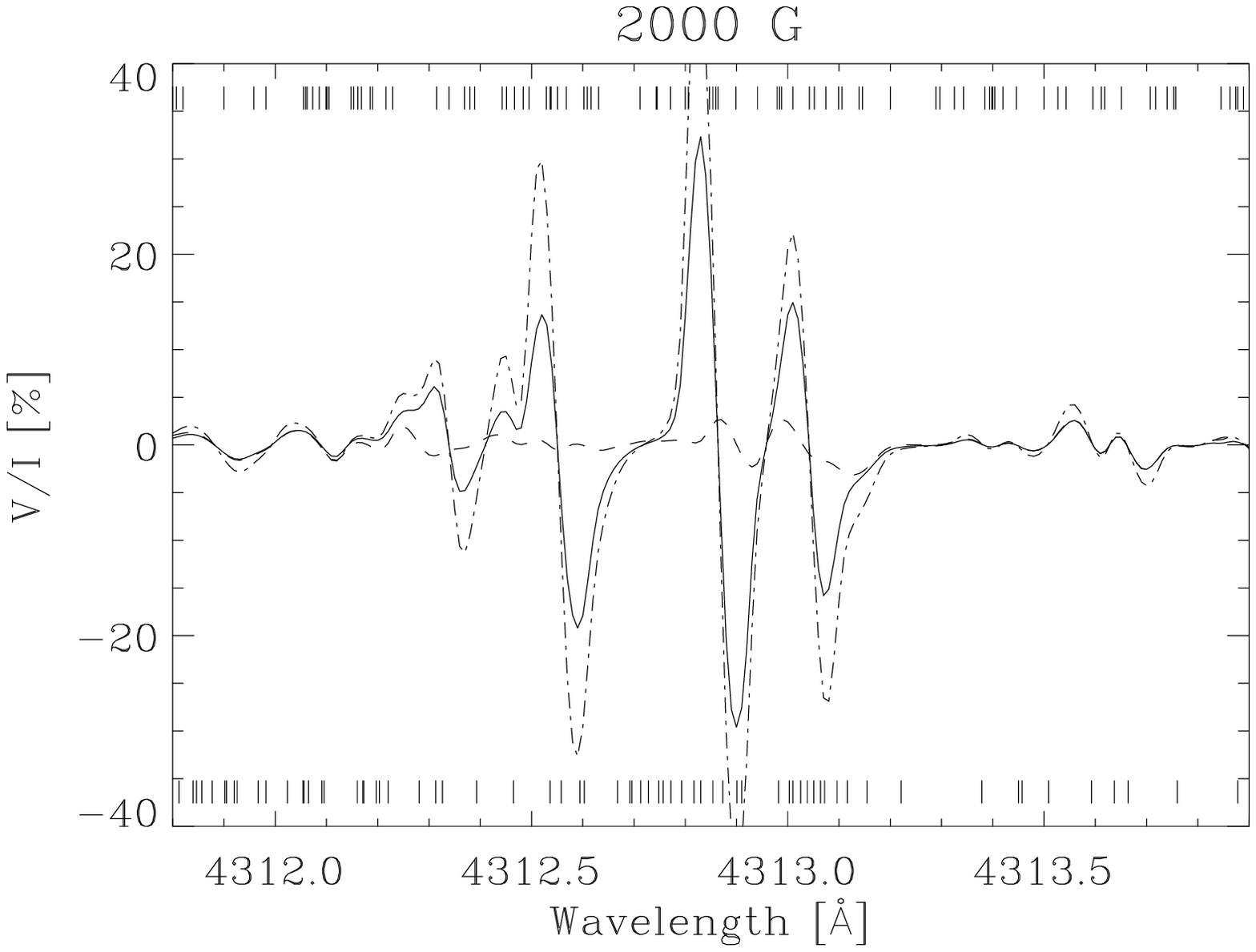}
\caption{The emergent $V/I$ profiles calculated
in the cool umbra model of
Collados et al
(1994) assuming a constant vertical magnetic
field of 2000 G and filling factor $f=1$.
The solid line represents the emergent fractional circular
polarization taking
into account both atomic and CH molecular
lines, while the dashed line represents the emergent $V/I$
including only CH lines. Both have been calculated considering the
effect of
a stray-light contamination of 4\% from the surrounding quiet Sun.
The dash-dotted line represents the calculated profiles without
any stray-light contamination.
The left panel shows the region around
4304 \AA\ in which we find the single-lobed
$V/I$ profile produced by CH alone. The
other circular polarization
signal is produced by an atomic line
of Fe {\sc i}. The right panel shows the region around 4313 \AA, where we can
find the $V/I$ profiles
at 4312 \AA\ and 4313.7 \AA\
which are produced exclusively by CH transitions. The
signals between both CH features are
produced mainly by
atomic lines. The strongest one is due to
Ti {\sc ii}. Each of the
tickmarks in the upper part of each panel
indicate the position of an atomic line, while
those in the lower part indicate the position of a CH line.}
\label{fig_syn}
\end{figure}

\begin{figure}
\epsscale{0.5}
\plotone{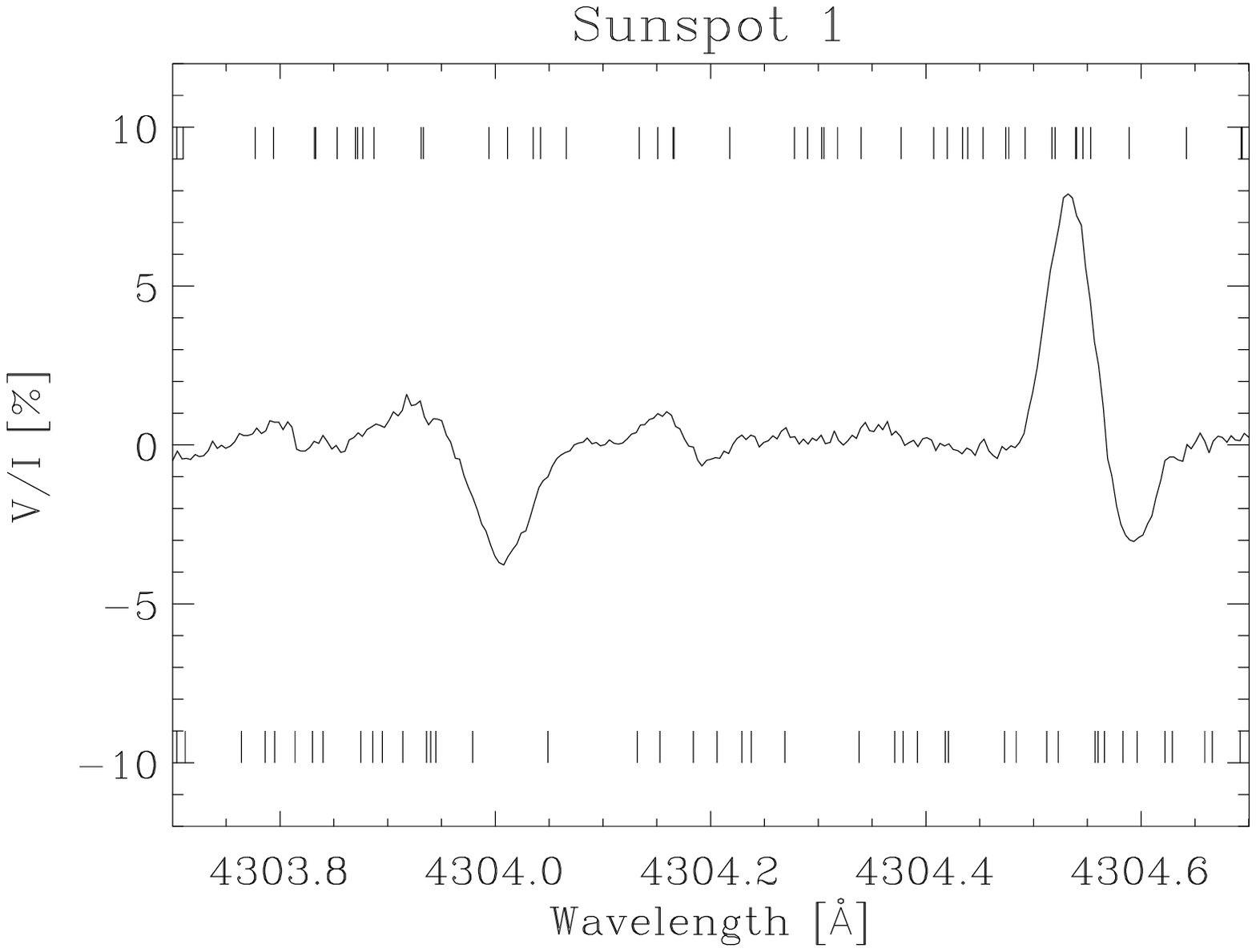}\\
\plotone{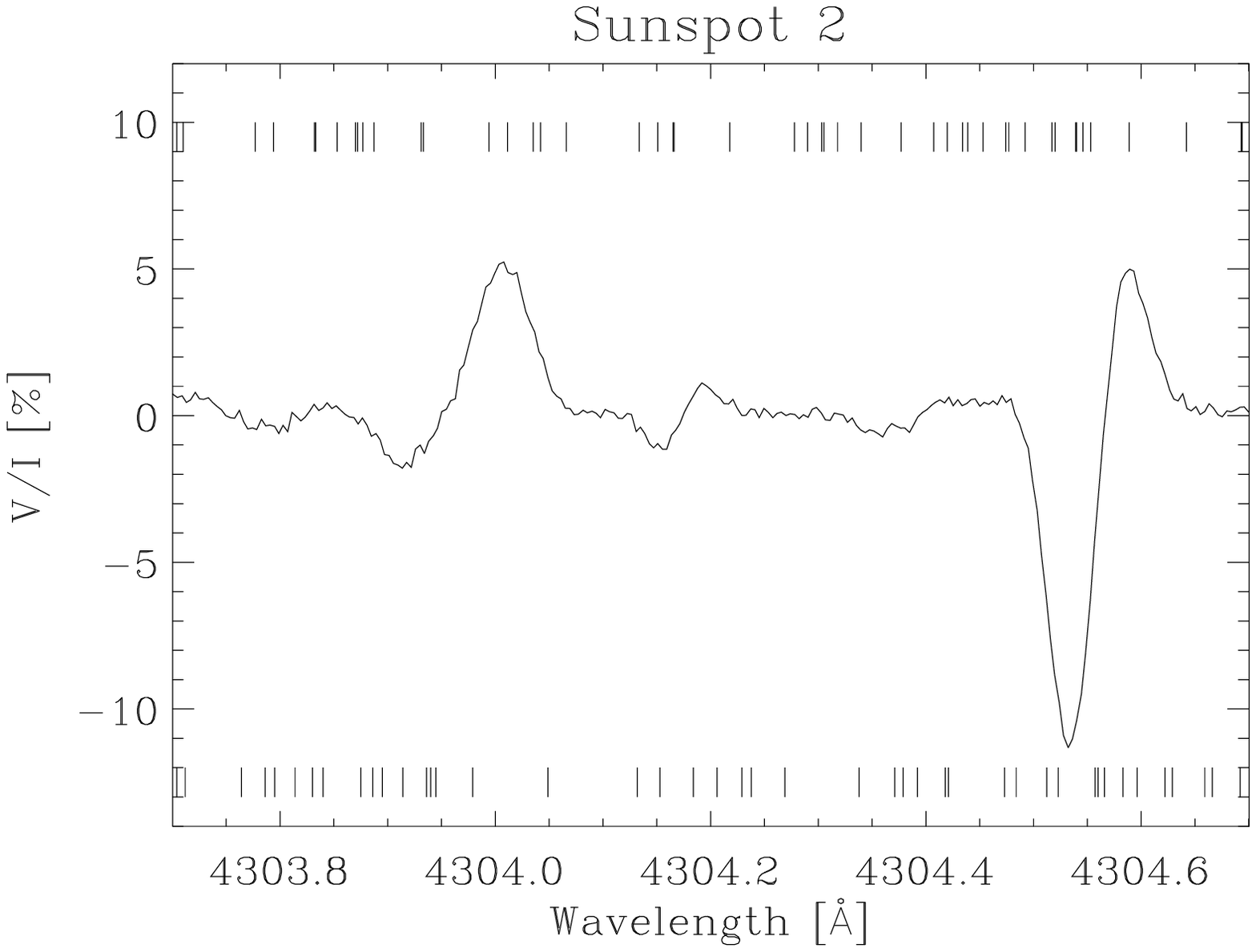}
\caption{The observed $V/I$ profiles in the two sunspots
with opposite magnetic polarity for
the 4304 \AA\ spectral region. Compare with Fig. \ref{fig_syn}
and note the good
agreement between the shapes of the calculated
and observed $V/I$ profiles.
On the other hand, the calculated $V/I$ amplitudes are
significantly larger than the observed ones. Each of the
tickmarks in the upper part of each panel
indicate the position of an atomic line, while
those in the lower part indicate the position of a CH line.}
\label{fig_obs_V4303}
\end{figure}

Figure \ref{fig_obs_V4303} shows examples of the observed Stokes $V/I$ profiles
around 4304 \AA, which is the wavelength position of the
CH lines that were predicted to produce a peculiar
Stokes $V/I$ profile dominated by its red lobe
(see Fig. \ref{fig_syn}). The top panel of Figure \ref{fig_obs_V4303} corresponds to the
sunspot with the positive polarity (i.e., with the magnetic field vector
pointing outwards), while the bottom
panel refers to the sunspot with the negative polarity. Each of the vertical marks in the lower part of the plot indicate
the position of a CH line while those in the upper part indicate the position of an atomic line. Note that each of the observed
signals is produced by a blend of many lines.
In full agreement
with our theoretical $V/I$ profiles of sunspot umbrae (see Fig. \ref{fig_syn})
the observed circular polarization at 4304 \AA\ is dominated by
a single-lobed Stokes $V/I$ profile whose sign changes when going
to the sunspot with the other magnetic polarity. The same occurs with the
Stokes $V/I$ profile at 4304.6 \AA, which is produced by atomic lines alone.

\begin{figure}
\epsscale{0.5}
\plotone{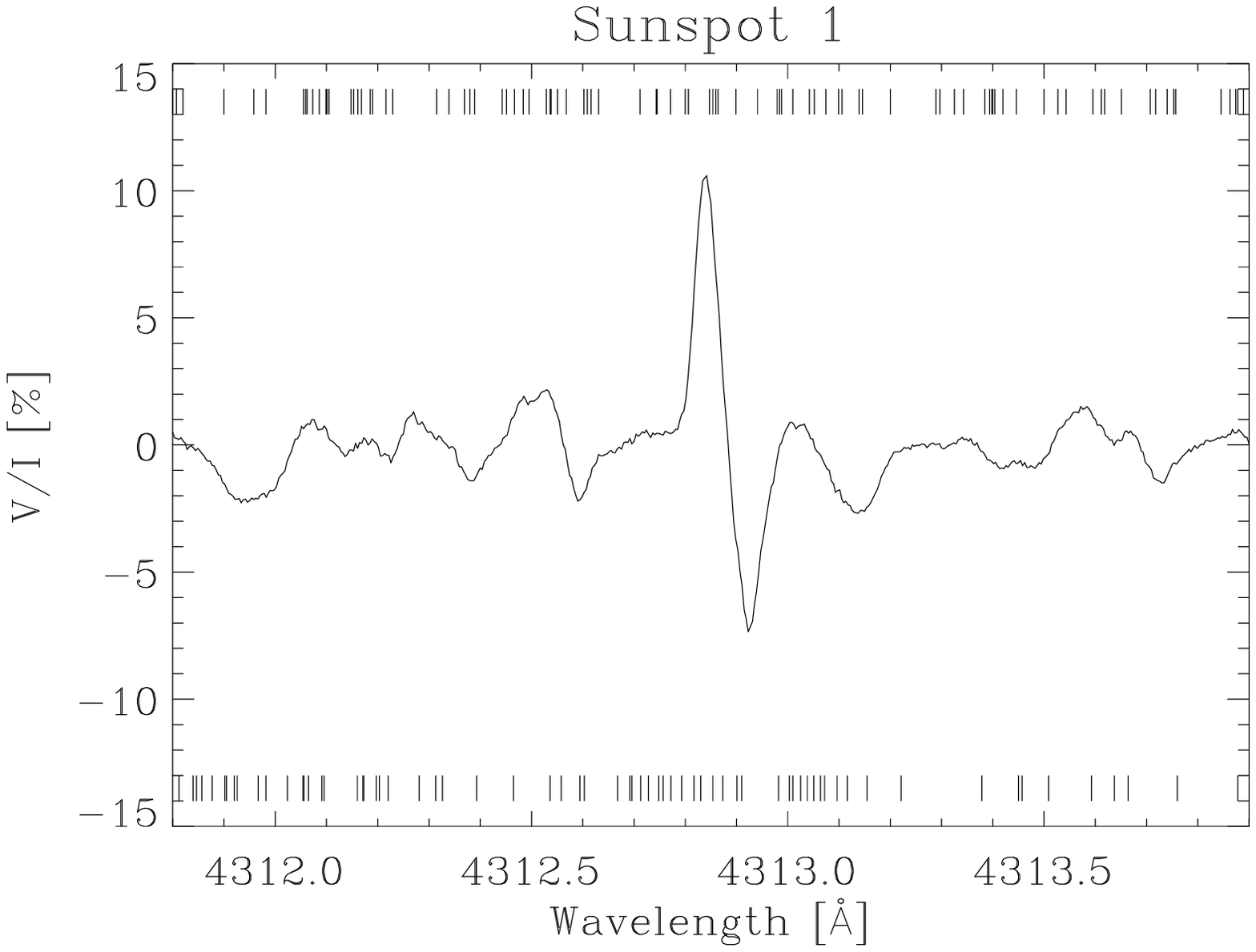}\\
\plotone{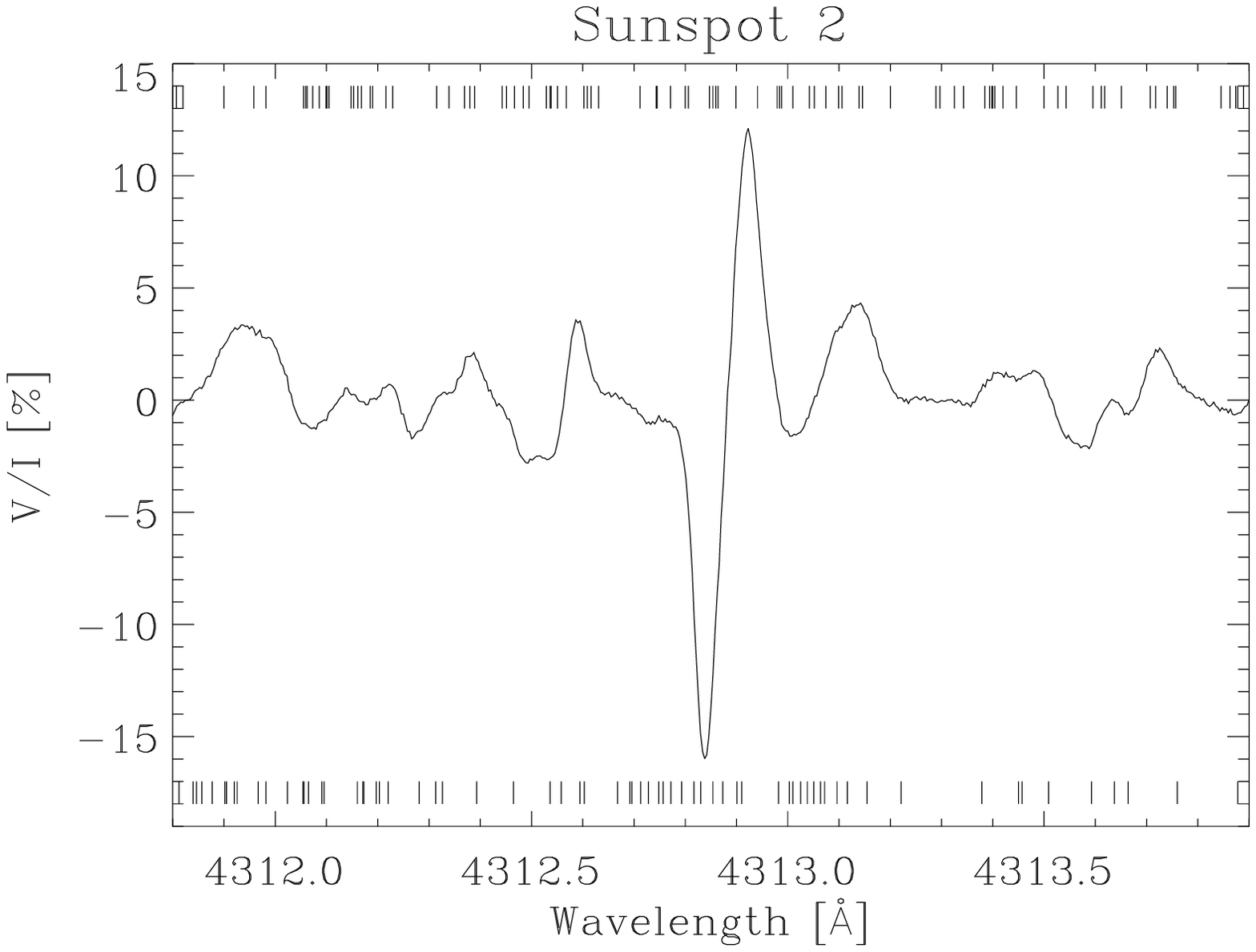}
\caption{The observed $V/I$ in the two sunspots
with opposite magnetic polarity for
the 4313 \AA\ spectral region. Compare with Fig. \ref{fig_syn}
and note the good
agreement between the shapes of the calculated
and observed $V/I$ profiles, especially concerning
the spectral features dominated by CH transitions.
Note that the curious shape of the $V/I$ profile produced by
CH lines at 4313.7 \AA\ is very similar to the theoretical
profile shown in Fig. \ref{fig_syn}.
The calculated $V/I$ amplitudes are again
significantly larger than the observed ones.
Each of the
tickmarks in the upper part of each panel
indicate the position of an atomic line, while
those in the lower part indicate the position of a CH line.}
\label{fig_obs_V4313}
\end{figure}

Figure \ref{fig_obs_V4313} shows the observed $V/I$ profiles for an
extra spectral region that also shows Stokes
profiles dominated by CH lines alone.
Note the polarity reversal between the two observed sunspots and the
good agreement with the shapes of the theoretical $V/I$ profiles
at 4312 \AA\ and 4313.7 \AA\ shown in Fig. \ref{fig_syn}. The Stokes $V/I$ profiles around
4313 \AA\ are mainly due to atomic lines.

It is also interesting to mention that the observed linear polarization
turns out to be very small at the above-mentioned wavelength locations
which show $V/I$ features dominated by CH lines. However, we have detected
sizable $Q/I$ signals (of the order of $1\%$) that are
produced by atomic lines alone. We find significant linear polarization
{\em in both umbrae and penumbrae}, with the signals in the penumbral regions being
slightly larger than in the umbral ones.
Figure \ref{fig_obs_Q4303} shows an example of the
observed fractional linear polarization in the penumbra of one of the observed
sunspots. We think that
the $Q/I$ feature at 4304 \AA\ is produced by the same CH lines
that are responsible of the observed $V/I$ profile shown in the left panel of Fig. \ref{fig_obs_V4303}.

A comparison with Fig. \ref{fig_syn} shows that
the agreement between the \emph{shapes}
of the calculated and the observed $V/I$ profiles
is remarkable.
However, the {\em amplitudes} of the
$V/I$ profiles calculated
in one-dimensional models of sunspot atmospheres
are significantly larger
than the observed ones when the theoretical modeling
is carried out assuming a magnetic filling
factor $f=1$ and a typical magnetic field strength of 2000 G,
even after taking into account the reduction in the $V/I$ amplitude
obtained when accounting for the stray-light contamination
mentioned in Section 2.
We consider this as an indication of the presence of horizontal inhomogeneities in the photospheric regions of
sunspot umbrae coexisting within the spatio-temporal resolution element of the observation. These inhomogeneities
might be associated with the multitude of umbral dots that are seen in high resolution images of sunspots. Interestingly, our finding of linear
polarization signals in the observed
sunspot umbrae, which were located very close to the
disk center during our observing period, could be indicating that the
suggested umbral dot
component has inclined magnetic fields, in agreement with the existing
magnetohydrodynamical models for the subsurface structure
of sunspots (see, e.g., the recent review by Socas-Navarro 2003).

\section{Conclusions}

\begin{figure}
\epsscale{1}
\plottwo{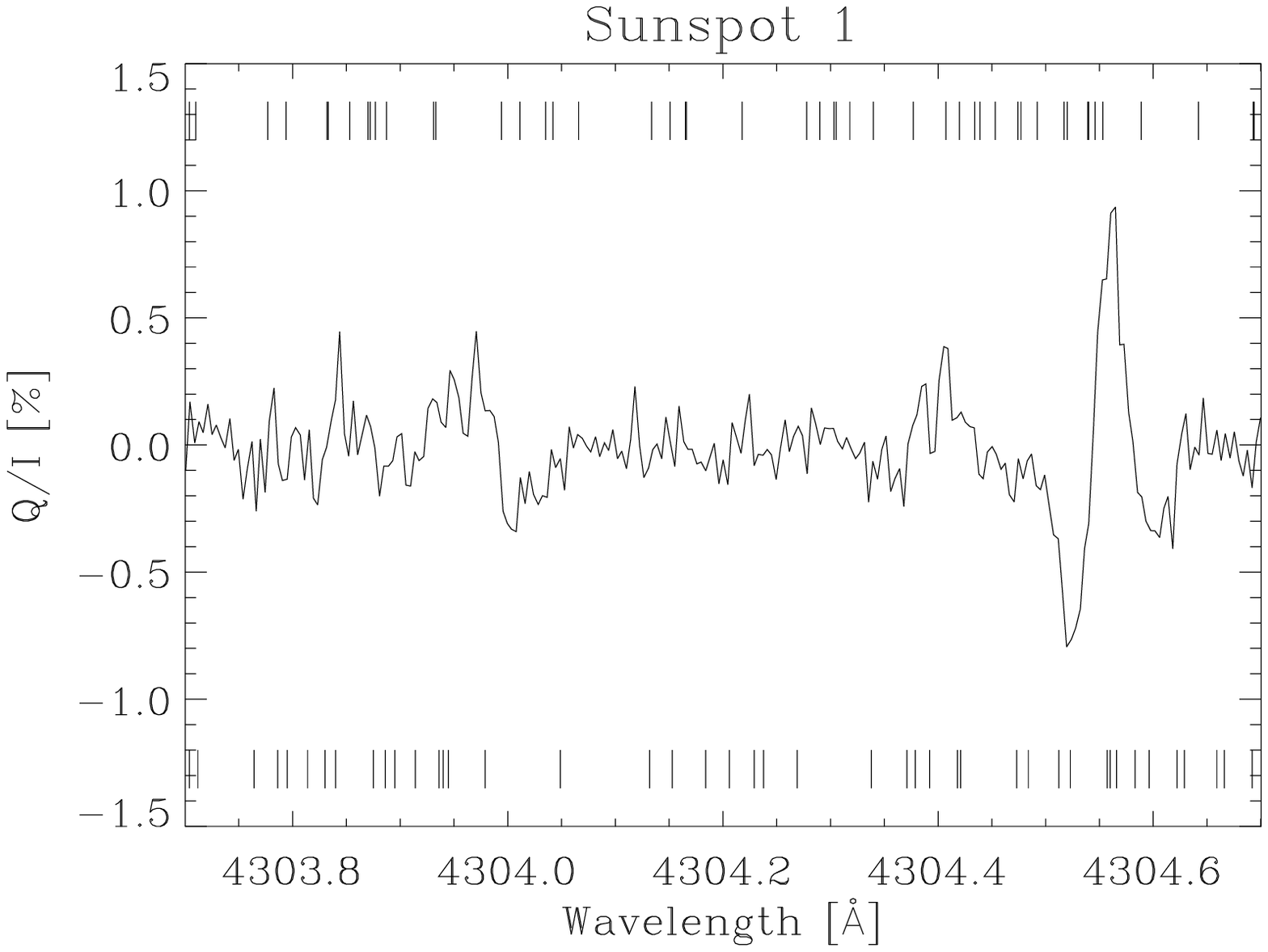}{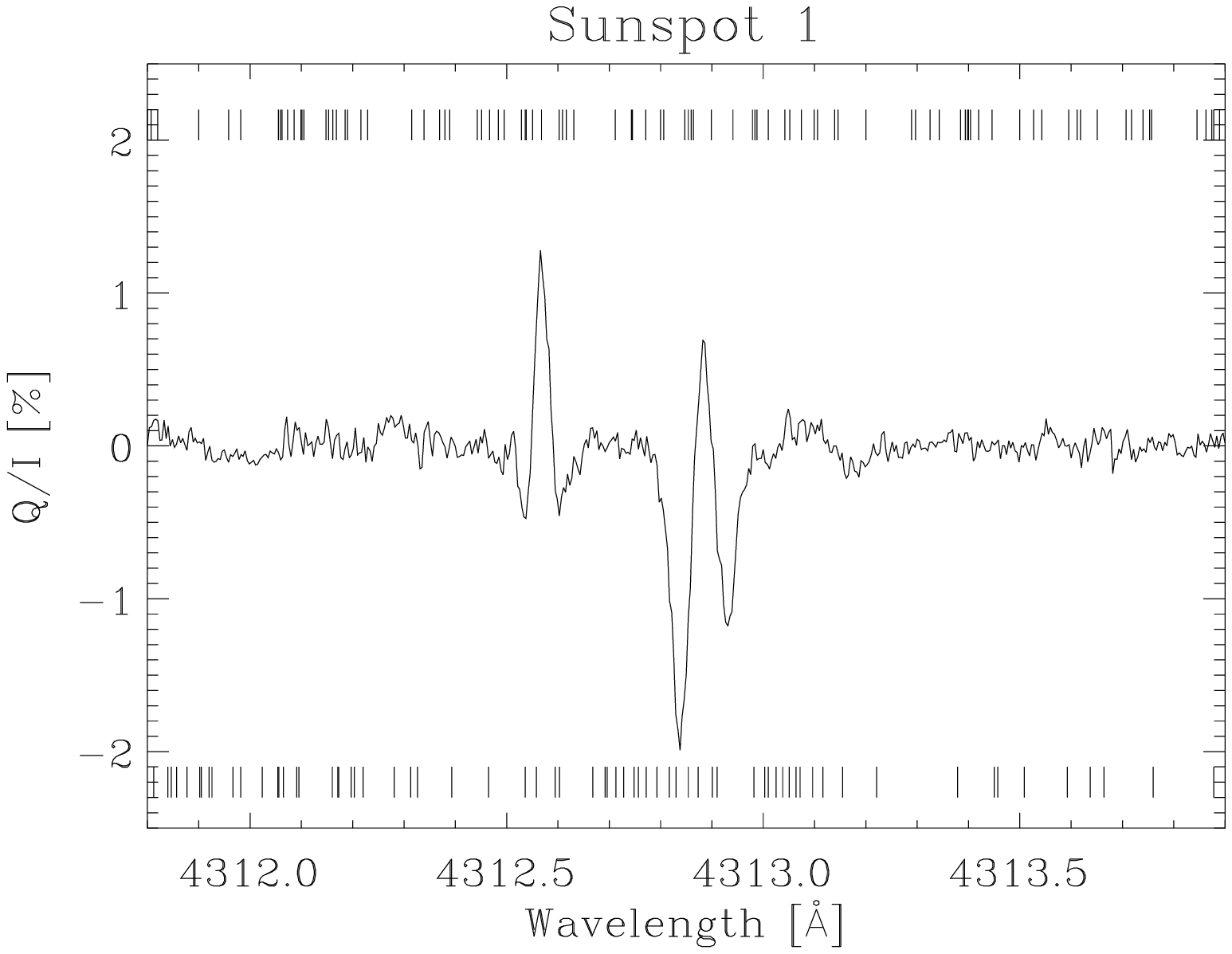}
\caption{Observed $Q/I$ profiles in the penumbra of
sunspot 1. Note that the strongest $Q/I$
signals are produced by the atomic
lines which show the strongest $V/I$ signals in Figs. \ref{fig_obs_V4303} and \ref{fig_obs_V4313}. The reference direction
for $Q>0$ is along the slit.}
\label{fig_obs_Q4303}
\end{figure}

The polarization profiles we have observed in sunspots
confirm our previous theoretical modeling of the longitudinal Zeeman effect
in the G band. There are at least three wavelength locations
which show measurable $V/I$ profiles that are produced by
CH lines alone. In the sunspot group we have observed
(located at $\mu=0.95$) the observed linear polarization at
the wavelength location of such CH lines was very small, but
not negligible (see, e.g., the $Q/I$ feature at 4304 \AA\ in Fig. \ref{fig_obs_Q4303}). However,
we have found sizable circular and linear polarization signals in many of the
atomic lines that are contained in the spectral region of the G band.
Such polarization signals in molecular and atomic lines contain valuable
information about the physical conditions in the solar atmosphere.
For instance, a preliminary comparison between observed and calculated
Stokes profiles
suggests that the magnetic and/or thermodynamical properties
in the photospheric regions
of sunspot umbrae are horizontally structured with a component that might be associated with umbral dots.

We think that the theoretical interpretation of
observations of the Zeeman effect in the G band
offers a new diagnostic window for exploring the thermal
and magnetic structuring of the solar photosphere. This type of investigations
could help us to choose among competing MHD models on the
subsurface structure of sunspot umbrae and/or to improve our physical understanding of the magnetic field strength in the bright points
of the `quiet' solar surface.
In a future paper we plan to
address the issue of the inversion of spectropolarimetric
observations in the G band, since this
promises also to be important for
improving our knowledge on solar and stellar magnetism.

\acknowledgments

J.T.B would like to express his gratitude to the Fondazione IRSOL
and to its President (Prof. Philippe Jetzer) for their kind invitation
to visit IRSOL. R.M.S. is also grateful to our IRSOL
colleagues for their hospitality.
We thank Jan Olof Stenflo and his ETH group for their invitation to
observe with the Z\"urich Imaging Polarimeter (ZIMPOL),
and to Achim Gandorfer  for his very helpful advice to achieve optimum
observations in the UV range.
This work has been partially supported by the Spanish Ministerio
de Ciencia y Tecnolog\'{\i}a through project AYA2001-1649.


\end{document}